\documentclass[12pt]{article}
\usepackage{amssymb}
\usepackage[,all]{xy}

\begin{document}
\newtheorem{prop}{}[section]
\newtheorem{defi}[prop]{}
\newtheorem{lemma}[prop]{}
\newtheorem{rema}[prop]{}
\def\parn{\par \noindent}
\def\fine{\hfill $\diamond$ \vskip 0.2cm \noindent}
\newcommand{\rref}[1]{(\ref{#1})}
\def\beq{\begin{equation}}
\def\feq{\end{equation}}
\def\parn{\par\noindent}
\def\lt{$LT$ }
\def\vain{\rightarrow}
\def\interi{{\textbf Z}}
\def\lag{Lagrange }
\def\kow{Kowalewski }
\def\bh{bH }
\def\qbh{qbH }
\hyphenation{di-men-sio-nal}
\newcommand{\CH}{{\cal H}}
\newcommand{\CP}{{\cal P}}
\newcommand{\CD}{{\cal D}}
\newcommand{\CA}{{\cal A}}
\newcommand{\CW}{{\cal W}}
\newcommand{\CZ}{{\cal Z}}
\newcommand{\CS}{{\cal S}}
\newcommand{\CU}{{\cal U}}
\newcommand{\CN}{{\cal N}}
\newcommand{\CV}{{\cal V}}
\newcommand{\CM}{{\cal M}}
\newcommand{\CQ}{{\cal Q}}
\newcommand{\CL}{{\cal L}}
\newcommand{\CK}{{\cal K}}
\newcommand{\del}{{\partial}}
\def\lag{Lagrange }
\def\kow{Kowalewski }
\def\bh{bH }
\def\qbh{qbH }
\makeatletter
\@addtoreset{equation}{section}
\renewcommand{\theequation}{\thesection.\arabic{equation}}
\makeatother
\begin{titlepage}
\begin{center}
{\huge Separation of variables in multi--Hamiltonian systems:
an application to the Lagrange top.}
\end{center}
\vspace{1truecm}
\begin{center}
{\large
Carlo Morosi${}^1$,~Giorgio Tondo${}^2$  } \\
\vspace{0.5truecm}
${}^1$ Dipartimento di Matematica, Politecnico di
Milano, \\ P.za L. da Vinci 32, I-20133 Milano, Italy \\
e--mail: carmor@mate.polimi.it \\
${}^2$ Dipartimento di Scienze Matematiche, Universit\`a di Trieste, \\
via A. Valerio 12/1, I-34127 Trieste, Italy \\
e--mail: tondo@univ.trieste.it \\
\end{center}
\vspace{1truecm}
\begin{abstract}
\noindent
Starting from the tri-Hamiltonian formulation of the Lagrange top in
a six-dimensional
phase space, we discuss the reduction of the vector field and of the
Poisson tensors. We show explicitly that, after the reduction on each
one of the  symplectic leaves, the vector field of the
Lagrange top is separable in the sense of Hamilton--Jacobi.
\end{abstract}
\vspace{0.2truecm} \noindent

\end{titlepage}
\section{Introduction}
\label{introd}
\parn
This paper completes the analysis of the Lagrange top ($LT$) as
a quasi-bi-Hamiltonian $(qbH)$ system started in \cite{mt2002},
to which we refer for more details. Summarizing, we showed in
\cite{mt2002} that the
tri--Hamiltonian structure of $LT$, defined on a six-dimensional
phase space by three compatible Poisson tensors
$P_0$, $P_1$, $P_2$, can be reduced onto a four-dimensional phase
space. When one tries to eliminate the
Casimirs of the Poisson tensors by fixing their values, one is faced
with a typical situation, occurring also for other
finite-dimensional integrable systems \cite{GT1, FMT, FMPZ}: to each one of the
symplectic leaves $S_0$, $S_1$, $S_2$ one can restrict only the
vector field  $X_L$ and the corresponding
Poisson tensor, but not the entire tri--Hamiltonian structure, which
is lost under restriction.
Nevertheless, the $LT$ vector field $X_L$, restricted
to the symplectic leaf $S_0$
of the Poisson tensor $P_0$, can be given a $qbH$ formulation, and 
this fact yields its separability in
the sense of Hamilton--Jacobi (HJ).
\parn
In this paper, we show that this property of the vector field $X_L$
is more general. In fact, the $qbH$ property can be recovered
also if the restriction is performed to any symplectic
leaf of the second Poisson tensor $P_1$; moreover, exploiting some 
properties of the $qbH$ model,
we can explicitly give the separation variables for the restriction 
of $X_L$ to any symplectic leaf
of the third Poisson tensor $P_2$. It is
remarkable that in both cases the separation variables are obtained by means of
non-symplectic maps.
\parn
The paper is organized as follows. In Section 2 and 3, we discuss 
some properties of a generic
tri--Hamiltonian structure with deformation and of the $qbH$ model, in view
of application to $LT$. In Sections 4 and 5 the tri--Hamiltonian structure
of $LT$, the deformation field and their reduction are briefly reviewed.
At last, the results of Section 3 are applied in Section 6 to show
that $X_L$ can be written in a separable form when its restriction is 
performed to any symplectic leaf
of the three Poisson tensors.
\section{Deformation of multi--Hamiltonian
structures and hereditary operators}
\label{sezione due}
Let us assume that:
\parn
{\bf i}) $(M, P_2, \tau)$ is a Poisson manifold with a deformation, i.e.,
$M$  is a differential manifold endowed with a Poisson tensor $P_2$
and with a vector field $\tau$ in such a way that the Lie
derivative $P_1=L_\tau(P_2)$ of $P_2$ w.r.t. $\tau$ is itself a Poisson
tensor. This assumption implies that $P_2-\lambda P_1$ is a Poisson pencil.
\parn
{\bf ii}) The Poisson tensor $P_1$ is exact w.r.t. $\tau$, i.e.,
$L^2_\tau(P_1)=0$. This implies that
$P_0=L_\tau(P_1)= L^2_{\tau}(P_2)$ is itself a Poisson tensor, compatible with
both $P_1$ and $P_2$.
\vskip0.1cm\noindent
So, under the above assumptions it follows that:
\begin{prop}
{\bf Lemma}
The manifold $(M, P_0,  P_1,  P_2)$ is a
tri--Hamiltonian manifold, i.e.,  the linear
combination $P_0 -\,\lambda P_1 \,-\mu P_2$ is itself a Poisson tensor
for every value of the constants coefficients $\lambda$ and $\mu$.
\end{prop}
{\bf Remark} As one can easily verify, if there is a
tri--Hamiltonian structure
($\tilde{P}_0$, $\tilde{P}_1$, $\tilde{P}_2$)
such that
$L_\tau ( \tilde{P}_0 )=0$, $L_{\tau} (\tilde{P}_1)= \alpha\,\tilde{P}_0$,
$L_{\tau} (\tilde{P}_2)= \beta \,\tilde{P}_1 + \gamma \tilde{P}_0$,
with $\alpha$, $\beta$ and $\gamma$ constant parameters, then
$ P_0=\tilde{P}_0$, $P_1=(1/ \alpha) \tilde{P}_1 + {(\gamma/
\alpha\beta)} \tilde{P}_0$ and
$P_2={(a / \alpha\beta)} \tilde{P}_2$ fulfil, for any given $a$, the
deformation relations
\beq
\label{deforM}
L_{\tau} (P_2)=a \,P_1~  \qquad  L_{\tau} (P_1)= P_0,  \qquad
L_{\tau} (P_0)=0~.
\feq
The following Lemma gives a sufficient condition for the projection of
the tri--Hamiltonian structure along a submersion.
\begin{prop} \label{propPi}
{\bf Lemma}
Let $\pi:M\mapsto M'$ be a surjective submersion onto a manifold
$M'$. If both $P_2$ and $\tau$ are
projectable onto $M'$, the whole tri--Hamiltonian
structure is preserved under the submersion $\pi$.
Denoting the reduced tensors as $P'_0$, $P'_1$, $P'_2$ and the
reduced vector field as
$\tau'$, it still holds that
\beq
\label{deforM'}
     L_{\tau'} (P'_2)=a \,P'_1~  \qquad  L_{\tau'} (P'_1)= P'_0,  \qquad
L_{\tau'} (P'_0)=0.
\feq
\end{prop}
As is known, if the reduced Poisson tensor $P'_0$ is kernel-free, so
that its inverse is symplectic, then on $M'$ the operator
$N=P'_1\,{P'_0}^{-1}$ is a hereditary operator, i.e., it has a vanishing
Nijenhuis torsion
\cite{MM}.  We now search for some conditions on the deformation $\tau$,
assuring that $N$ acts as a recursion operator (in a direction
opposite to the one of the Lie derivatives
w.r.t. $\tau$) for the reduced
tri--Hamiltonian structure on $M'$, mapping also
$P'_1$ to $P'_2$ (possibly, up to a constant factor):
$NP'_1= \lambda \, P'_2$ for some constant $\lambda$; to this
purpose, the following result can be used.
\vskip 0.2cm
\noindent
\begin{prop}
\label{lemmadue}
{\bf{Lemma}} Given a vector field $\tau$ on $M$ and its reduction
$\tau'$ on $M'$, let
us consider the  equation $ L_{\tau'}(Q')=0 $ and look for a solution
$Q'$ which is a (2,0) skew-symmetric tensor. If $P'_0$ is the
general solution (up to a constant factor): 
$L_{\tau'}(Q')=0\,\Rightarrow\,Q'= \alpha\, P'_0$,
then the tensor $N:=P'_1\, {P'_0}^{-1}$ defined on the  manifold
$M'$ is such that
\beq
\label{qqq}
N\,P'_1= {2\over a}\, P'_2 + \beta\, P'_0 \qquad (\beta=constant)~.
\feq
\end{prop}
{\bf{Proof}} On account of \rref{deforM'} and the previous assumptions, the
equations $L_{\tau'} (Q')= P'_0$ and $L_{\tau'} (Q')=
P'_1$  have the solutions  $Q'= P'_1 + \alpha P'_0$ and $Q'=
(1/a)\,P'_2 + \alpha P'_0$, respectively;
since it is also $L_{\tau'}(N)=I$, we have
$$
{1\over 2-a}\,L_{\tau'} (NP'_1-P'_2)= P'_1\qquad\Rightarrow\qquad
{1\over 2-a}\,(NP'_1-P'_2)={1\over a} P'_2 + \alpha P'_0
$$
yielding \rref{qqq} with $\beta=(2-a)\,\alpha $.
\begin{prop}
\label{lemmauno}
{\bf{Lemma}} Let a tri--Hamiltonian structure ($P'_0$, $P'_1$,
$P'_2$) be given, with $\tau'$
fulfilling \rref{deforM'} and a recursion operator $N$ such that
$N\,P'_0=P'_1$ and $N\,P'_1 = \lambda \,P'_2 + \mu\,P'_0$~.
Then the tensors $Q_0=P'_0$, $Q_1=P'_1$, $Q_2=P'_2+ (\mu/\lambda)\,
P'_0$ are such that the deformation relations
\rref{deforM'} still hold and $N\,Q_0=Q_1$, $N\,Q_1= \lambda \,Q_2$~.
\end{prop}
{\bf{Proof}} A trivial computation.
\vskip 0.1cm\noindent
On account of the above results, given a tri--Hamiltonian structure
$(P_0$, $P_1$, $P_2)$ with a deformation $\tau$
fulfilling \rref{deforM}, if the deformation and the tri--Hamiltonian
structure are preserved under the submersion $\pi$,
then on the reduced manifold there is also (possibly after a
rescaling) a recursion structure defined by $N$~.
\section{Some properties of the quasi-bi-Hamiltonian model}
\label{reminder}
The $qbH$ model was introduced in \cite{cab,bro}
and developed in \cite{mt,tm}.
\parn
Let $Q_0$, $Q_1$ be two compatible Poisson tensors;
a vector field $X$ admits a $qbH$ formulation if there
are three functions $\rho$, $H$, $K$ such that
\beq
X= {Q}_0 ~dH= {1\over \rho} ~{Q}_1 ~dK~.
\label{ro}
\feq
If $M$ is even-dimensional, $\mbox{dim}~M=2n$, the $qbH$ formulation
is said to be of maximal
rank if $Q_0$, $Q_1$ are
non degenerate at each point $m\in M$ and the associated tensor
$N=Q_1~ Q_{0}^{-1}$
(with vanishing Nijenhuis torsion) has $n$ distinct
eigenvalues $\lambda_1(m),..., \lambda_n(m)$; the $qbH$
formulation is said to be of Pfaffian type if $\rho=\prod_{i=1}^n \lambda_i$.
For a $qbH$ structure of maximal rank, one can introduce a
Darboux-Nijenhuis chart
$(\lambda_i, \mu_i)$~$(i=1,2,...,n)$ such that $Q_0$, $Q_1$ and $N$
take the canonical form 
\beq
Q_0=~\left(
\begin{array}{cc}
0 & I \\
-I& 0 \\
\end{array}
\right),
\quad
Q_1=~\left(
\begin{array}{cc}
0 & \Lambda \\
-\Lambda & 0 \\
\end{array}
\right),\qquad
\quad
N=~\left(
\begin{array}{cc}
\Lambda &0 \\
0&\Lambda  \\
\end{array}
\right)
\label{diag}
\feq
where $I$ is the $n\times n$ unit matrix and $\Lambda=\mbox{diag}
(\lambda_1,...,\lambda_n)$\,\cite{mm}.
\parn
\begin{prop}
\label{hkgen}
{\bf Proposition \cite{mt}} In a  Darboux-Nijenhuis chart, the
general solution of Eq.\rref{ro} for the
Pfaffian case is given by the functions
\beq
H=\sum_{i=1}^n {f_i \over \Delta_i}, \qquad
K=\sum_{i=1}^n {\rho\over \lambda_i}~{f_i \over \Delta_i}
\label{gen}
\feq
where $\Delta_i= \prod_{j\neq i} (\lambda_i-\lambda_j)$~ and
$f_i$ are arbitrary  functions, depending at most
on the pair $(\lambda_i, \mu_i)$.
Moreover, the $HJ$ equations for both $H$ and $K$ are
separable in the chart $(\lambda, \mu)$.
\end{prop}
\noindent
>From now on, functions of the form (\ref{gen}) in a given chart will 
be said to have a
{\it normal} form. The above  Proposition yields straightforwardly the
following.
\begin{prop}
\label{rem2}
{\bf Corollary} Let $X=Q_0~dH$ be a Hamiltonian vector field, $Q_0$ and $H$
taking the  form \rref{diag}, \rref{gen}; then there exist $Q_1$
and $K$ of the form \rref{diag} and \rref{gen}, respectively,
such that $X= (1/\rho) \,Q_1 dK$.
\parn
Viceversa, let $X= (1/\rho) \,Q_1~dK$, with $\rho=\prod_{i=1}^n
\lambda_i$,  $Q_1$ and $K$ of the  form
\rref{diag} and \rref{gen}; then there exist $Q_0$
and $H$ of the form \rref{diag} and \rref{gen}, respectively, such that
it is also $X= Q_0 dH$.
\end{prop}
In view of applications to $LT$, let us consider in more detail 
a four-dimensional manifold $(n=2)$;
in this case, we have some more general conditions assuring that
a Hamiltonian vector field is separable.
\begin{prop}
\label{hkgen1}
{\bf Proposition} On a four-dimensional manifold, let $ X=Q_0~d\tilde{H} $ be a
Hamiltonian vector field, with $Q_0$ of the canonical form
\rref{diag} in a chart $(x;y)$. Let the Hamiltonian $\tilde{H}$ be a linear
combination of two functions $\hat{H}$, $\hat{K}$ possessing  the 
normal form \rref{gen}, i.e.,
\beq
\label{hhh}
\tilde{H}(x;y)= \alpha \,\hat{H}(x;y) + \beta\, \hat{K}(x;y) \qquad
(\alpha, \beta=\mbox{const.}) ~,
\feq
$$\hat{H}(x;y)={1\over x_1-x_2} \left(\hat{f}_1(x_1, y_1)-
\hat{f}_2(x_2, y_2)\right)~,
$$
$$
\hat{K}(x;y)={1\over x_1-x_2} \left(x_2 \hat{f}_1(x_1, y_1)- x_1
\hat{f}_2(x_2, y_2)\right)~.
$$
Then the map $\Phi_0:(x;y)\mapsto (\lambda;\mu)$
\begin{equation} \label{eqPhi0}
\lambda_i={\beta \over  \alpha+\beta\,x_i}~,\quad \mu_i=-{1\over 
\beta^2}\,( \alpha+\beta\, x_i)^2\, y_i
\qquad(i=1,2)
\end{equation}
is symplectic for $Q_0$ (i.e., $Q_0$ is preserved under
$\Phi_0$); $\tilde{H}$ is transformed
under $\Phi_0$ into a function $H$ of the normal form \rref{gen}, with
\beq
\label{film}
f_i(\lambda_i , \mu_i)=-\beta\,\lambda_i\,\hat{f}_i 
({1\over\lambda_i} -{\alpha\over \beta}~,
-\lambda_i^2\mu_i)
\qquad (i=1,2)~.
\feq
So, the vector field $X$ admits a $qbH$ formulation and the Hamiltonian
$\tilde{H}$ is separable in the chart  $(\lambda;\mu)$.
\end{prop}
{\textbf{Proof}}
By straightforward computations one checks that the map $ \Phi_0$ is
symplectic for $Q_0$ and that the Hamiltonian $\tilde{H}$ takes the form
\beq
\label{form}
\tilde{H} \Big(x(\lambda;\mu);y(\lambda;\mu)\Big)= {1\over
\lambda_1-\lambda_2} \Big(f_1(\lambda_1, \mu_1)
-f_2(\lambda_2, \mu_2)\Big)~,
\feq
with $f_i(\lambda_i , \mu_i)$ given by \rref{film}.
On account of Corollary \ref{rem2}, the vector field
$X= Q_0 dH$ admits also the $qH$ formulation $X=1/\rho~ Q_1 dK$;
the separability of $\tilde{H}$ in the chart $(\lambda; \mu)$ follows
from Proposition \ref{hkgen}.
\begin{prop}
\label{RemSoVxy}
{\bf Corollary} $\tilde{H}$ is separable also in the chart $(x;y)$;
the corresponding $HJ$ equation $\tilde{H}(x;\partial W / \partial 
x)=\tilde{h}$ has the complete solution
$ W = W_1+ W_2$, $W_1$ and $W_2$ fulfilling the Jacobi
separation equations
\cite{Sk}
\beq
\hat{f}_i(x_i, W'_i(x_i))= x_i \hat{h} -\hat{k},\quad 
\alpha\,\hat{h}+\beta\,\hat{k}=\tilde{h}\qquad(i=1,2) ~.
\label{eqSk}
\feq
\end{prop}
Indeed,  the map $\Phi_0$ is a {\it separated}  map \cite{Be}, i.e., it maps
separated coordinates into separated ones.
So, taking into account the form \rref{form} of the function $\tilde{H}$, it
is easily checked that the $HJ$ equation has a complete solution
$ W(x_1, x_2; \hat{h}, \hat{k})=$ $W_1(x_1; \hat{h}, \hat{k})+
W_2(x_2; \hat{h}, \hat{k})$, with $\alpha\,\hat{h} +\beta \,\hat{k}=
\tilde{h}~$ and $W_1$, $W_2$ fulfilling the Jacobi separation equations
(\ref{eqSk}) for the $HJ$ equations $\hat{H}(x;\partial
W/\partial x)=\hat{h}$, ~
$\hat{K}(x;\partial W/\partial x)=\hat{k}$.
\begin{prop}
\label{fiuno}
{\bf{Proposition}} On a four-dimensional manifold, let $X=Q_1
\,d\tilde{H} $ be a
Hamiltonian vector field, with $Q_1$ of the  form \rref{diag} in a
chart $(x; y)$; let the Hamiltonian $\tilde{H}$
be a linear combination of two functions $\hat{H}$, $\hat{K}$ with 
the normal form \rref{gen}, i.e.,
$$
\tilde{H}(x;y)= \alpha \,\hat{H}(x;y) + \beta\,\hat{K}(x;y) \qquad
(\alpha, \beta=\mbox{const.}) ~,
$$
$$\hat{H}(x;y)={1\over x_1-x_2} \left(\hat{f}_1(x_1, y_1)-
\hat{f}_2(x_2, y_2)\right)~,
$$
$$
\hat{K}(x;y)={1\over x_1-x_2} \left(x_2 \hat{f}_1(x_1, y_1)- x_1
\hat{f}_2(x_2, y_2)\right)~.
$$
Then the map $\Phi_1:(x;y)\mapsto (\lambda,\mu)$ given by
\begin{equation}
\label{eqPhi1}
\lambda_i={1\over \alpha+ \beta\,x_i}~,
\quad
\mu_i=-{1\over \beta}\,(\alpha+ \beta \,x_i)^2\,{y_i\over x_i} \qquad
(i=1,2)
\end{equation}
is a Darboux map for $Q_1$ (i.e., $Q_1$ is mapped to $Q_0$ under 
$\Phi_1$); $\tilde{H}$ is transformed under $\Phi_1$
into a function  $H$ of the normal form \rref{gen}, with
\beq
\label{ffff}
f_i(\lambda_i, \mu_i)= -\beta \,\hat{f}_i \Big ( {1\over
\beta}({1\over\lambda_i}-\alpha), -({1\over\lambda_i}-\alpha)
\,\lambda_i^2 \,\mu_i \Big)~~~(i=1,2).
\feq
So, the vector field $X$ admits a $qbH$ formulation and the 
Hamiltonian $\tilde{H}$ is
separable in the chart $(\lambda; \mu)$.
\end{prop}
{\bf{Proof}} A straightforward computation allows one to check that 
$\Phi_1$ is a Darboux map for $Q_1$ and that $\tilde{H}$ is mapped 
into a function $H$ of normal form, with $f_i$ given by (\ref{ffff}). 
Corollary \ref{rem2}
and Proposition \ref{hkgen} assure that $\tilde{H}$ is separable in 
the chart $(\lambda, \mu)$.
\begin{prop}
\label{fidue}
{\bf{Proposition}} On a four-dimensional manifold, let $ X=Q_2\, 
d\tilde{G}$ be a
Hamiltonian vector field, with $Q_2$ of the form \rref{diag} in a 
chart $(x;y)$ and
$$
\tilde{G}=\alpha\,\hat{K}+ \beta \hat{H}^2~,
$$
$\hat{H}$ and $\hat{K}$ being in the normal form \rref{gen}.
Then the map $\Phi_2:(x;y)\mapsto (\lambda;\mu)$
\beq
\label{eqPhi2}
\lambda_i= -{1\over x_i}~,\qquad \mu_i=y_i
\quad(i=1,2)
\feq
is a Darboux map for $Q_2$ (i.e., it maps $Q_2$ into $Q_0$); 
$\tilde{G}$ is mapped under $\Phi_2$ into the function
\beq
G= \alpha \,H+\beta\, K^2
\feq
with $H$ and $K$ in the normal form \rref{gen} and
\beq
\label{ffdue}
f_i(\lambda_i, \mu_i)= -\lambda_i \, \hat{f}_i (-{1\over\lambda_i}, \mu_i )
\qquad\qquad(i=1,2)~.
\feq
The function $\tilde{G}$ is separable in the chart $(\lambda; \mu)$.
\end{prop}
{\bf{Proof}} It is straightforward to verify that $\Phi_2$ is a Darboux map
for $Q_2$ and that $\tilde{G}$ is mapped into
$G$.  Let us consider the $HJ$ equation
$G(\lambda; \partial W/\partial \lambda)=g$ for $G$; one can easily verify that
it is
$W=W_1+W_2$ with
$W_1$ and $W_2$ solutions of the separation equations
\beq
f_i(\lambda_i, W'_i(\lambda_i))=\lambda_i\,h-k\qquad
(\alpha\,h+ \beta\,k^2=g)\qquad(i=1,2)~.
\feq
\section{The tri--Hamiltonian structure of the Lagrange top}
\label{rev}
In the comoving frame, whose axes are the
principal inertia axes of the top, with fixed point $O$, the \lt is
parametrized by the pair $m=(\omega;\gamma)$,
where $\omega = (\omega_1, \omega_2, \omega_3)^T~$  and $\gamma =
(\gamma_1, \gamma_2,
\gamma_3)^T~$ are the angular velocity and the vertical unit vector,
respectively. If $\mu$ is the mass of the top, $g$ the
acceleration of gravity, $J=\mbox{diag}(A, A, cA)$ the principal inertia matrix
$(c\neq 1)$ and $G=(0,0,a)^T$ the center of mass,
normalisations are chosen so that $\mu a g = A$.
\parn
The Euler-Poisson equations $dL_o/dt=M_o$ and $d\gamma/dt=0$ take the form
$dm /dt=X_L(m)$, where $X_L$ is given by
$$
X_L(m)= \Big(
-(c-1) \omega_2 \omega_3-\gamma_2, (c-1) \omega_3 \omega_1+\gamma_1, 0;
\gamma_2 \omega_3- \gamma_3 \omega_2,
\gamma_3 \omega_1- \gamma_1 \omega_3,
\gamma_1 \omega_2- \gamma_2 \omega_1\Big)^T~.
$$
The \lt vector field $X_L$ can be given a tri--Hamiltonian formulation
$$
X_L= P_0 dh_0= P_1 dh_1= P_2 dh_2~;
$$
written in matrix block form, the compatible Poisson tensors are:
$$
{P}_0= \left(
\begin{array}{cc}
0  & B \\
B & C \\
\end{array}
\right)~,
\qquad
{P}_1= \left(
\begin{array}{cc}
-B  & 0 \\
0 & \Gamma \\
\end{array}
\right)~,
\qquad{P}_2= \left(
\begin{array}{cc}
T  & R \\
-R^T & 0 \\
\end{array}
\right)~,
$$
$$
B= \left(
\begin{array}{ccc}
0  & -1 & 0 \\
1 & 0 & 0 \\
0 & 0& 0 \\
\end{array}
\right),
\
C= \left(
\begin{array}{ccc}
0  & c ~\omega_3 & -\omega_2 \\
- c ~\omega_3 & 0 & \omega_1 \\
\omega_2 & -\omega_1 & 0 \\
\end{array}
\right),
\
\Gamma = \left(
\begin{array}{ccc}
0  & \gamma_3 & -\gamma_2 \\
-\gamma_3 & 0 & \gamma_1 \\
\gamma_2 & -\gamma_1 & 0 \\
\end{array}
\right),
$$
$$
T= \left(
\begin{array}{ccc}
0  & - c~ \omega_3 & \omega_2/ c \\
c~ \omega_3 & 0 & - \omega_1 / c \\
- \omega_2 / c & \omega_1/ c & 0 \\
\end{array}
\right)~,
\quad
R = \left(
\begin{array}{ccc}
0  & -\gamma_3 & \gamma_2 \\
\gamma_3 & 0 & -\gamma_1 \\
-\gamma_2/ c & \gamma_1 /c & 0 \\
\end{array} \right) ~.
$$
The Hamiltonian functions are
$$
h_0={1\over 2} F_4+ (c-1) F_1 F_3,~~~
h_1= {1\over 2} c (c-1)  F_{1}^3- F_3- (c-1) F_1 F_2,~~~
h_2 = F_2
$$
\beq
\label{ham}
F_1=\omega_3,\qquad\qquad  F_2={1\over 2} (\omega_{1}^2+\omega_{2}^2+
c~ \omega_{3}^2)
-\gamma_3,
\feq
$$
F_3= \omega_1\gamma_1+\omega_2\gamma_2+ c~ \omega_3\gamma_3~,\qquad
F_4= \gamma_{1}^2+\gamma_{2}^2+ \gamma_{3}^2~.
$$
The functions $(F_1, F_2)$ are Casimirs of $P_0$,
$(F_1, F_4)$ of $P_1$ and $(F_3, F_4)$ of $P_2$.
\parn
The Hamiltonian formulation of \lt w.r.t. $P_2$ is
classical (see, e.g.,\cite{gav});
the \bh formulation w.r.t. $(P_0, P_2)$ was introduced in \cite{rat}
in  the semidirect product $\mathfrak{so}(3)\times \mathfrak{so}(3)$, and
was later recovered in \cite{med} in an algebraic-geometric  setting;
the tri--Hamiltonian formulation w.r.t. $(P_0, P_1, P_2)$ was
constructed in \cite{mag}, by a suitable reduction of the Lie-Poisson
pencil defined in the direct sum of three copies of $\mathfrak{so}(3)$.
\parn
As shown in \cite{mag}, the tri--Hamiltonian structure of $LT$ admits
the deformation $L_{\tau}(P_2) = 2 P_1$, $L_\tau(P_1) = P_0$, 
$L_\tau(P_0) = 0$,
where $\tau$ is given, in the chart $(\omega; \gamma)$, by
$\tau=(0,0,-2/c; \omega_1,\omega_2,c\, \omega_3)^T$; on the contrary,
a recursion operator $N$ relating the Poisson
tensors does not exist in $M$.
\parn
The Poisson pencils $P_1-\lambda P_0$, $P_2-\lambda P_1$,
$P_2-\lambda P_0$ are three Poisson pencils of Gelfand--Zakharevich
(GZ)
type: more precisely, they belong to the class of complete torsionless
GZ systems of rank $2$ \cite{GZ}.  Each one of them has two
polynomial Casimir functions, whose coefficients form two   Lenard
chains for each pencil which can be constructed by means of the 
deformation field
$\tau$. Graphically, the Lenard chains of
$(P_0, P_1)$ can be represented in the following way:
$$
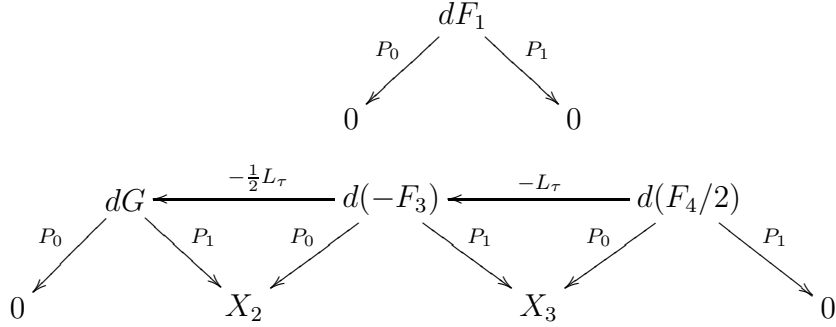
\xymatrix{
&dF_1
\ar[dl]_{P_0} \ar[dr]^{P_1}&\\
0&&0
}
$$
$$
\xymatrix{
&dG \ar[dl]_{P_0} \ar[dr] ^{P_1}
& &d(-F_3)\ar[ll]_{-{1\over 2} L_{\tau}} \ar[dl]_{P_0} \ar[dr]^{P_1}
& &d(F_4/2)\ar[ll]_{- L_{\tau}}\ar[dl]_{P_0} \ar[dr]^{P_1}& \\
0& &X_2& &X_3& &0&
}
$$
where $G=F_2+c(c-1)F_1^2/2$.
The Lenard chains of $(P_1, P_2)$ are:
$$
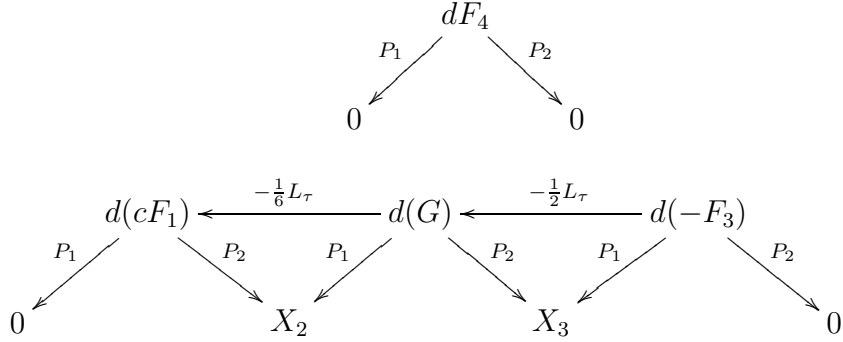
\xymatrix{
&dF_4
\ar[dl]_{P_1} \ar[dr]^{P_2}&\\
0&&0
}
$$
$$
\xymatrix{
&d(c F_1) \ar[dl]_{P_1} \ar[dr] ^{P_2}
& &d(G) \ar[ll]_{-{1\over 6} L_{\tau}}\ar[dl]_{P_1} \ar[dr]^{P_2}
& &d(-F_3)\ar[ll]_{-{1\over 2} L_{\tau}}\ar[dl]_{P_1} \ar[dr]^{P_2}& \\
0& &X_2& &X_3& &0&
}
$$
  So, we can state that $P_1-\lambda P_0$
and $P_2-\lambda P_1$ are Poisson pencils of rank $2$ and
type $(1,5)$. Finally, the Lenard chains of $(P_0, P_2)$ are:
$$
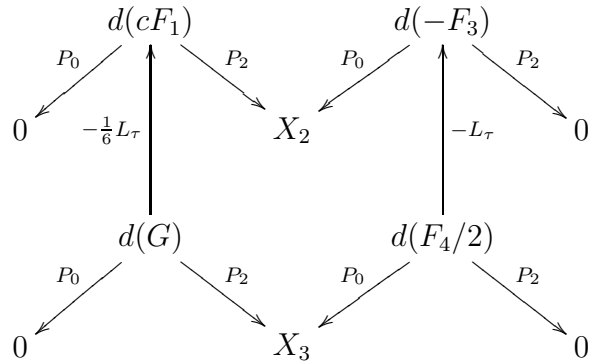
\xymatrix{
&d(c F_1)
\ar[dl]_{P_0} \ar[dr]^{P_2}
& &d(-F_3) \ar[dl]_{P_0} \ar[dr]^{P_2}&\\
0& &X_2& &0& \\
&d(G)\ar[uu]^ {-{1\over 6} L_{\tau}}\ar[dl]_{P_0} \ar[dr] ^{P_2}
& &d(F_4/2) \ar [uu]_{- L_{\tau}}\ar[dl]_{P_0} \ar[dr]^{P_2}& \\
0& &X_3&  &0&
}
$$
implying that $P_2-\lambda P_0$ is Poisson pencil of rank $2$ and
type $(3,3)$.
\section{The reduction of the tri--Hamiltonian and deformation structures}
The tri--Hamiltonian structure ($P_0, P_1, P_2$) of $LT$ and the
deformation field $\tau$ admit a reduction on a
four-dimensional manifold $M'$ (see \cite{mt2002} for an
interpretation of this process
in terms of the Marsden-Ratiu reduction theorem).
\parn
Let $M'$ be a four-dimensional manifold parametrized by a chart
$(x;y)=(x_1, x_2; y_1, y_2)$, and let
$\pi:M\mapsto M'=M/\pi,\,:(\omega; \gamma)\mapsto (x;y)$ be the
surjective submersion given by:
$$
x_{1,2}= - {1\over 2} \,(c\,\omega_3 -i\,\omega_2) \mp{1\over 2}\,
\sqrt{(c\,\omega_3-i\,\omega_2)^2+ 4(\gamma_3-i\,\gamma_2)}
$$
$$y_{1,2}=- \gamma_1 -{1\over 2}
\,\omega_1(c\,\omega_3-i\,\omega_2)\mp {1\over 2}\,\omega_1
\sqrt{(c\,\omega_3-i\,\omega_2)^2+4(\gamma_3 -i\,\gamma_2
)}~.
$$
A straightforward calculation allows one to conclude the following.
\begin{prop}
\label{riduz}
{\bf{Proposition}} The Poisson tensor $P_0$ and the deformation
field $\tau$ can be reduced under $\pi$: the projected tensor fields 
take the form
\beq
\label{canonical}
P'_0=-i~\left(
\begin{array}{cc}
        0 & I \\
-I& 0 \\
\end{array}
\right)~\qquad
\tau'=(1,1,0,0)^T
\feq
where $I$ is the $2\times 2$ unit matrix.
On account of Proposition \ref{propPi}, also $P_1$ and $P_2$ are projectable:
the reduced tensors $P'_1$, $P'_2$ take the  form
\beq
P'_1=-i~\left(
\begin{array}{cc}
0 & \cal X \\
-\cal X & 0 \\
\end{array}
\right)~\qquad
P'_2=-i~\left(
\begin{array}{cc}
0 & {\cal X} ^2 \\
-{\cal X }^2& 0 \\
\end{array}
\right)~,
\feq
where $ {\cal X} =\mbox{diag} \,(x_1, x_2)$.
\end{prop}
Moreover, the deformation relations are maintained under $\pi$.
Since $P'_0$ is clearly kernel-free, one has a
torsionless tensor
$N= P'_1 \,{P'_0}^{-1}$; it can be easily checked that $N\,P'_0=P'_1$,
$N\, P'_1= P'_2$. So, we are just in the situation discussed in Lemma 
\ref{lemmadue},
with $a=2$, $\beta=0$~.
\section{The reduction of the vector field $X_L$ on the symplectic leaves}
\label{reddue}
In this section we consider the reduction of $LT$ on the symplectic
leaves $S_i$ of the Poisson tensors
$P_i$ $(i=0,1,2)$. Each $S_i$ is  a four-dimensional submanifold of $M$,  being
characterized as a level set of two Casimirs functions of $P_i$ . On account of
Eq.\rref{ham}, the symplectic leaves are defined as
\beq
S_0=\{ m\in M~\vert~ \omega_3={C_1},~~
\omega_{1}^2+\omega_{2}^2+ c \omega_{3}^2-2\gamma_3=2\,C_2 \}~,
\label{leave0}
\feq
$$
S_1=\{ m\in M~\vert~ \omega_3= C_1,
~~\gamma_{1}^2+\gamma_{2}^2+  \gamma_{3}^2= C_4\}~,
$$
$$
S_2=\{ m\in M~\vert~  \omega_1\gamma_1+\omega_2\gamma_2+ c
\omega_3\gamma_3 = C_3~,~~\gamma_{1}^2+\gamma_{2}^2+  \gamma_{3}^2 = C_4\}
$$
where $C_1$, $C_2$, $C_3$ and $C_4$ are fixed values of the Casimirs
$F_1$, $F_2$, $F_3$ and $F_4$, respectively.
As these integrals of motion are in involution w.r.t. each $P_i$,    a
level set, on
$S_i$, of the other two integrals of motion is a leaf $\Lambda_i$ of a
Lagrangian foliation of $S_i$ . Let us note that for any $m  \in M$,
the Lagrangian leaves $\Lambda_0$, $\Lambda_1$, $\Lambda_2$, passing
through $m$, coincide. Moreover, using the Marsden-Ratiu reduction
theorem, one can prove the following result.
\begin{prop}
\label{diffeom}
{\bf{Proposition \cite{mt}}} The symplectic leaves $S_0, S_1, S_2$ are
(locally) diffeomorphic to the four-di\-men\-sio\-nal
manifold $M'= M/ \pi$.
\end{prop}
Explicitly, let $(x_1, x_2; y_1, y_2)$ be a chart of $M'$; one can verify
that the symplectic leaves admit the following parametrizations.
\begin{prop}
\label{param}
{\bf{Lemma
}} A generic symplectic leaf $S_0$ is parametrized by the mapping
$\Psi_0:M'\mapsto M:$ $(x;y) \mapsto (\omega;\gamma)$ given by
$$
\omega_1={y_2-y_1\over x_2-x_1}, \quad \omega_2= -i ~(x_1+x_2 +
c\,C_1), \quad \omega_3= C_1
$$
$$
\gamma_1={x_1\,y_2- x_2\,y_1\over x_2-x_1}~,~~~
\gamma_2= {i\over 2} ~[x_1^2+x_2^2+g_1(x;y)]~,~~~
\gamma_3= -{1\over 2} [(x_1+x_2)^2 + g_1(x;y)]
$$
$$
g_1(x; y)= -{(y_2-y_1)^2\over (x_2-x_1)^2}+
2c\,C_1(x_1+x_2)+ 2C_2+ c(c-1)\,C_1^2~.
$$
A symplectic leaf $S_1$ is parametrized by the mapping
$\Psi_1$ given by
$$
\omega_1={y_2-y_1\over x_2-x_1}, \quad \omega_2= -i~(x_1+x_2 +
c\,C_1), \quad \omega_3= C_1
$$
$$
\gamma_1={x_1\,y_2- x_2\,y_1\over x_2-x_1}~,~~~
\gamma_2= -{i\over 2} ~[x_1\,x_2 -g_2(x;y)]~,~~~
\gamma_3= -{1\over 2} [x_1\,x_2 + g_2(x;y)]
$$
$$
g_2(x; y)= -{(x_2\,y_1-x_1\,y_2)^2\over x_1\,x_2\, (x_2-x_1)^2}
+{C_4\over x_1\,x_2}~.
$$
A symplectic leaf $S_2$ is parametrized by the mapping
$\Psi_2$ given by
$$
\omega_1={y_2-y_1\over x_2-x_1}~,~~~ \omega_2= -{i\over 2}
~[x_1+x_2 + g_3(x;y )]~,~~~
\omega_3=  -{1\over 2 c} ~[x_1+x_2 - g_3(x;y )]
$$
$$
\gamma_1={x_1\,y_2- x_2\,y_1\over x_2-x_1}~,~~~
\gamma_2= -{i\over 2} ~[x_1\,x_2 -g_2(x;y)]~,~~~
\gamma_3= -{1\over 2} [x_1\,x_2 + g_2
(x;y)]
$$
$$
g_3(x;y)= {x_1^2\,y_2^2- x_2^2\,y_1^2\over
x_1^2\,x_2^2\,(x_2-x_1)}-\,{2\,C_3\over x_1 x_2} + {x_1+x_2\over
x_1^2 \,x_2^2} \,C_4~.
$$
\end{prop}
By means of these parametrizations, we can show that the $LT$ admits
a qbH formulation on each one of the symplectic leaves.
The following three Propositions are easily proved by straightforward
computations.
\begin{prop}
\label{so}
{\bf Proposition} The vector field $X_L$, restricted to $S_0$,
takes the form
$$
X_L=P'_0 \,d\tilde{H},
$$
with $P'_0$ given by \rref{canonical}. The Hamiltonian
$\tilde{H}={h_0}\big\vert_{S_0}$ is given by
\beq
\label{hh0}
\tilde{H}(x; y)= (c-1)\,C_1~ \hat{H}(x; y)+ \hat{K}(x; y)
\feq
$$
\hat{H}(x; y)=F_3\vert_{S_0}={1\over x_1-x_2}\left(\hat{f}(x_1,
y_1)-\hat{f}(x_2,
y_2)\right)~,
$$
$$
\hat{K}(x; y)=\frac{F_4}{2}\vert_{S_0}={1\over x_1-x_2}\left(x_2
\hat{f}(x_1, y_1)-x_1
\hat{f}(x_2, y_2)\right)~,
$$
\beq
\label{ff0}
\hat{f}(\xi, \eta)= -{1\over 2}~ \eta^2 + {1\over 2}~ \xi^4 + c\,C_1 \xi^3 +
\Big(C_2+ {1\over 2}\, c(c-1)\,C_1^2\Big )\,\xi^2~.
\feq
\end{prop}
So, we are in the situation discussed in Proposition \ref{hkgen1},
with $\alpha= (c-1)\,C_1$ and $\beta=1$; this allows us to
conclude that in the chart $(\lambda, \mu)$ given by (\ref{eqPhi0})
$X_L\big\vert_{S_0}$ is a separable qbH vector field.
\vskip 0.1cm\noindent
{\bf{Remark}} The vector field $X_L\big\vert_{S_0}$ is separable also
in the chart $(x;y)$. Indeed, from Eq.\rref{eqSk} of Corollary
\ref{RemSoVxy} and from the expression \rref{ff0} of
$\hat{f}$ it follows that the solution $W$ of the $HJ$ equation for
$\tilde{H}$ is $W=W_1+W_2$, where
$W_1$ and $W_2$ can be computed  solving by quadratures the ODEs
obtained replacing $y_i$ by ${\partial W /\partial x_i}$ in the equation
\begin{equation} \label{eqscurve}
y_i^2=  x_i^4+ 2c\,C_1x_i^3 + [2 C_2 + c(c-1)\,C_1^2]\, x_i^2
- 2 \hat{h} x_i + 2 \hat{k}~\quad(i=1,2)~;
\label{wx}
\end{equation}
here, $\tilde{h}=(c-1)\,C_1 \,\hat{h}+\hat{k}$ and $\tilde{h}$, $\hat{h}$, $\hat{k}$ 
are the values of $\tilde{H}$, $\hat{H}$ and $\hat{K}$,
respectively, on a Lagrangian leaf
$\Lambda_0(C_3,C_4)$. Now, we are able to make contact with the Sklyanin method
of SoV
\cite{Sk}. Indeed, comparing (\ref{eqscurve}) with the spectral curve coming
from the  Lax pair
\cite{rat}, it immediately follows that the separation variables
$(x;y)$ satisfy the
equation of the spectral curve restricted to $\Lambda_0(C_3,C_4)$.
\vskip 0.1cm\noindent
Now, let us consider the reduction on a generic symplectic leaf $S_1$.
\begin{prop}
\label{suno}
{\bf Proposition} The vector field $X_L$, restricted to $S_1$,
takes the form
$$
X_L=P'_1 \,d\tilde{H}~
$$
with $P'_1$ given by \rref{canonical}. The Hamiltonian
$\tilde{H}=h_1\big\vert_{S_1}$ can be written as
\beq
\label{hh1}
\tilde{H}(x; y)= (c-1)\,C_1 \,\hat{H}(x; y)+ \hat{K}(x; y) ~,
\feq
$$
\hat{H}(x; y)=-G\vert_{S_1}={1\over x_1-x_2}\left(\hat{f}(x_1, 
y_1)-\hat{f}(x_2,
y_2)\right)~,
$$
$$
\hat{K}(x; y)={1\over x_1-x_2}\left(x_2 \hat{f}(x_1, y_1)-x_1
\hat{f}(x_2, y_2)\right)~,
$$
$$
\hat{f}(\xi, \eta)= -{1\over 2\,\xi}~ \eta^2 + {1\over 2}~ \xi^3 +
c\,C_1 \,\xi^2 +{C_4\over 2\,\xi} - {1\over 2} c (c-1) (c-2)\, C_1^3~.
$$
\end{prop}
We are in the situation discussed in Proposition \ref{fiuno},
with $\alpha=(c-1)\,C_1$ and $\beta=1$; so, we can conclude that the chart
$(\lambda; \mu)$ given by (\ref{eqPhi1}) provides the separation
variables for $\tilde{H}$.
  A solution of the HJ
equation for $\tilde{H}$
is given by:
\begin{equation}
W(\lambda_1,\lambda_2; \tilde{h},\hat{k})= \int^{\lambda_1}
\sqrt{\hat{\varphi}_1(\xi)}~ d\xi+
\int^{\lambda_2} \sqrt{\hat{\varphi}_1(\xi)}
~ d\xi
\end{equation}
$$
\hat{\varphi}_1(\xi)= {1\over\xi^4}\left(
({1\over\xi}-\alpha)^2+2c\,C_1({1\over\xi}-\alpha )-
\frac{\alpha(c-2)C_1^2-2\tilde{h}+{2\hat{k}\over\xi}}{{1\over\xi}
-\alpha}+\frac{C_4}{({1\over\xi}-\alpha)^2}\right)
\label{w1}
$$
where $\tilde{h}$ and $\hat{k}$ are   the values of $\tilde{H}$ and 
$\hat{H}$ on
a Lagrangian leaf $\Lambda_1( C_2,C_3)$.
\parn
At last, passing to the reduction on the symplectic leaf $S_2$, one
has the following
result.
\begin{prop}
\label{sdue}
{\bf Proposition} The vector field $X_L$, restricted to $S_2$,
takes the form
$$
X_L=P'_2 \,d\tilde{G}~
$$
with $P'_2$ given by \rref{canonical}. The Hamiltonian
$\tilde{G}=h_2\big\vert_{ S_2}$ can be written as
\beq
\tilde{G}=-{1\over 2 c}\,(c-1)\,\hat{H}^2 +  \hat{K}
\feq
$$
\label{hh2}
\hat{H}(x; y)=-cF_1\vert_{S_2}={1\over x_1-x_2}\Big(x_2 \hat{f}(x_1, y_1)-x_1
\hat{f}(x_2, y_2)\Big)
$$
$$
\hat{K}(x; y)=G\vert_{S_2}={1\over x_1-x_2}\left(x_2 \hat{f}(x_1, y_1)-x_1
\hat{f}(x_2, y_2)\right)~,
$$
$$
\hat{f}(\xi, \eta)=  -{1\over 2}\left({\eta^2\over \xi^2} - \xi^2 +2
{C_3\over
\xi} - {C_4\over \xi^2}\right ).
$$
\end{prop}
So, we are just in the situation of Proposition \ref{fidue}, with 
$\alpha= -(c-1)/2c$ and $\beta=1$;
also in this case, the vector field $X_L\big\vert_{S_2}$ is separable in the chart
$(\lambda; \mu)$ given by (\ref{eqPhi2}). A solution of the HJ
equation for $\hat{K}$
is given by:
\begin{equation}
W(\lambda_1,\lambda_2; h, k)= \int^{\lambda_1} \sqrt{\hat{\varphi}_2(\xi)}~
d\xi+
\int^{\lambda_2} \sqrt{\hat{\varphi}_2(\xi)}
~ d\xi
\end{equation}
$$
\hat{\varphi}_2(\xi)= \Big(
C_4+\frac{2C_3}{\xi}+\frac{2\hat{k}}{\xi^2}-\frac{2\hat{h}}{\xi^3}+\frac{1}{\xi^4}\Big)
\label{w2}
$$
where $\hat{h}$ and $\hat{k}$  are the values of $\hat{H}$ and $\hat{K}$   on a
Lagrangian leaf $\Lambda_2( C_1,C_2)$.
{\bf Remark} In contrast with what happens on $S_0$, let us observe that on $S_1$ and $S_2$ the LT vector
field $X_L$  is separable in the chart $(\lambda; \mu)$ but not
in the chart $(x;y)$, in which it does not admit a  
Hamiltonian formulation w.r.t. the canonical Poisson tensor $Q_0$.
\vskip 0.2cm
\noindent
{\bf Acknowledgments.}
This work was partially supported by the
research project {\textsl{Geometry of Integrable Systems}} of
M.I.U.R. and by G.N.F.M. of I.N.D.A.M.


\end{document}